# Rapid change of superconductivity and electron-phonon coupling through critical doping in Bi2212

Y. He[1,2,†], M. Hashimoto[3,†], D. Song[4,°], S.-D. Chen[1,2], J. He[1,2], I. M. Vishik[1,†], B. Moritz[1,2], D.-H. Lee[5], N. Nagaosa[6], J. Zaanen[7], T. P. Devereaux[1,2], Y. Yoshida[4], H. Eisaki[4], D. H. Lu[3], Z.-X. Shen[1,2,*]

**Affiliations:**

[1]Geballe Laboratory for Advanced Materials, Departments of Physics and Applied Physics, Stanford University, Stanford, California 94305, USA

[2]SIMES, SLAC National Accelerator Laboratory, Menlo Park, California 94025, USA.

[3]Stanford Synchrotron Radiation Lightsource, SLAC National Accelerator Laboratory, Menlo Park, California 94025, USA

[4]National Institute of Advanced Industrial Science and Technology, Tsukuba 305-8568, Japan

[5]Department of Physics, University of California, Berkeley, CA 94720–7300, USA

[6]Quantum-Phase Electronics Center, Department of Applied Physics, University of Tokyo, Tokyo 113-8656, Japan

[7]Instituut-Lorentz for Theoretical Physics, Leiden University, Leiden, Netherlands.

*Correspondence to: zxshen@stanford.edu.

†Present address: Department of Physics, University of California, Davis, California 95616, USA.

°Present address: Department of Physics and Astronomy, Seoul National University, Seoul 08826, Korea

**Abstract:** Electron-boson coupling plays a key role in superconductivity for many systems. However, in copper-based high-temperature ($T_c$) superconductors, its relation to superconductivity remains controversial despite strong spectroscopic fingerprints. Here we use angle-resolved photoemission spectroscopy to find a striking correlation between the superconducting gap and the bosonic coupling strength near the Brillouin zone boundary in $Bi_2Sr_2CaCu_2O_{8+\delta}$. The bosonic coupling strength rapidly increases from the overdoped Fermi-liquid regime to the optimally doped strange metal, concomitant with the quadrupled superconducting gap and the doubled gap-to-$T_c$ ratio across the pseudogap boundary. This synchronized lattice and electronic response suggests that the effects of electronic interaction and the electron-phonon coupling re-enforce each other in a positive feedback loop upon entering the strange metal regime, which in turn drives a stronger superconductivity.

**Main Text:** The phase diagram of cuprate high temperature superconductors hosts a number of complex orders, types of fluctuations and interactions (*1 - 4*). In the non-Fermi liquid strange metal regime, a hierarchy of microscopic interactions are intimately at play but not fully understood (*1, 2, 4*). Although the experimental evidence for *d*-wave superconductivity (*5 - 7*) naturally points to an electron-electron interaction based pairing mechanism (*8 - 12*), the omnipresent charge order (*3*) points to the role of electron-phonon coupling (EPC), especially in a new context of enhanced EPC by electronic correlation (*13, 14*) and multichannel boosted superconductivity (*15 - 17*). Although there have been reports of EPC imprinting on the electronic structure of many cuprate superconductors (*18 - 21*), little evidence directly correlates EPC with the intertwined orders in the phase diagram (*1 - 2*). Focusing on the overdoped side in $Bi_2Sr_2CaCu_2O_{8+\delta}$ (Bi-2212), we find via angle-resolved photoemission spectroscopy (ARPES) a set of striking effects rapidly crossing over from the overdoped Fermi-liquid regime to the optimally doped strange metal, closely associated with the putative pseudogap quantum critical point.

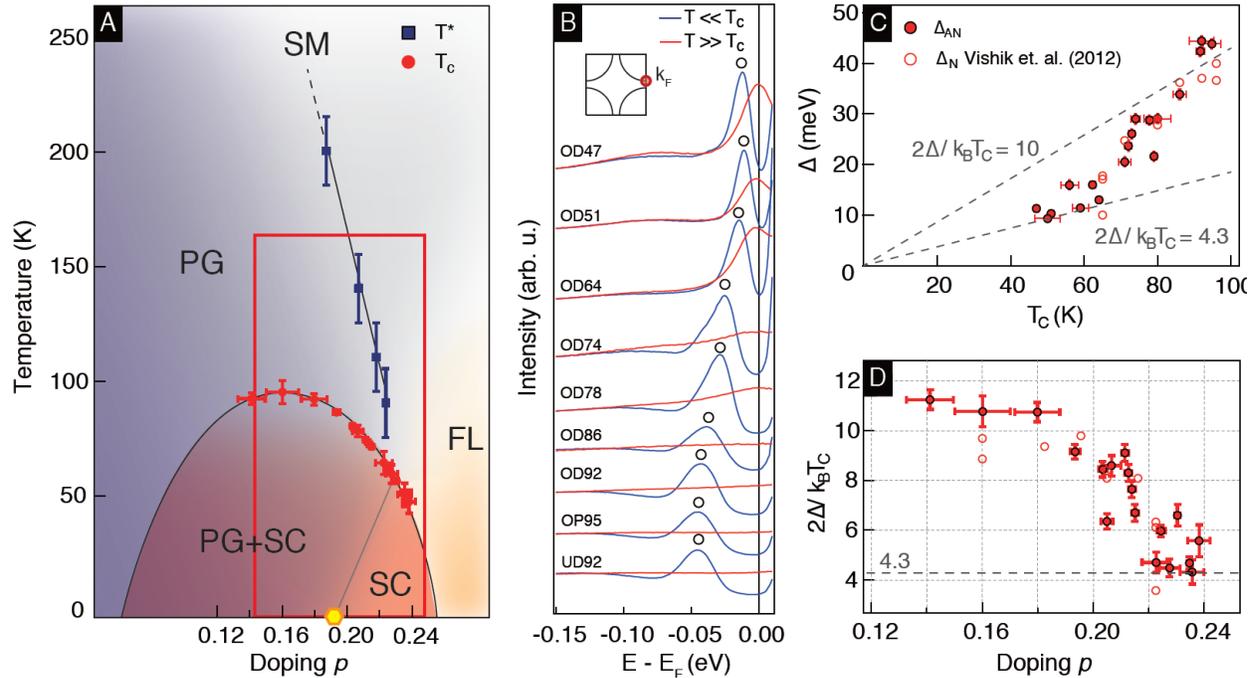

**Fig. 1. Superconductivity rapidly deviates from the weakly coupled *d*-wave BCS limit with underdoping**. (A) The schematic phase diagram for Bi-2212 reproduced from ref (*22*), with an emphasis on the doping range of interest (boxed in red). The blue squares are the antinodal gap opening temperature T*, and the red squares are the superconducting transition temperature $T_c$ (Fig. S1). PG - pseudogap, SC - superconductivity, SM – strange metal, FL - Fermi liquid. (B) The FD-EDCs taken at the antinodal $k_F$ for a series of doping. Red curves refer to T ~ T* for *p* < 0.22 and T > $T_c$ for *p* > 0.22, and blue curves refer to T << $T_c$ for all dopings (*24*). The black circles denote the superconducting quasiparticle peak position. The inset schematically illustrates the momenta of the listed spectra. (C) The antinodal gap $\Delta_{AN}$ (solid circle) and the nodal gap $\Delta_N$ (open circle) (*22*) plotted against the sample $T_c$. The two dashed lines mark the gap-to-$T_c$ ratios of 4.3

(the weakly coupled *d*-wave BCS value) and 10. (D) The gap-to-$T_c$ ratio shows a rapid deviation from 4.3 (horizontal black dashed line) at $p < 0.22$.

To understand the doping dependence of the superconducting character, we first investigate the energy gap. Figure 1A depicts the recently proposed phase diagram with an emphasis around the pseudogap phase boundary (red boxed area) (*22, 23*). In this doping range, the system is marked by a strange metal normal state characterized by spectral incoherence and linear resistivity at high temperatures (*2*), followed by intertwined states with pseudogap and superconductivity at low temperatures (*23*). Figure 1B plots the Fermi-Dirac function divided energy distribution curves (FD-EDCs) at the antinodal Fermi momentum $k_F$ from hole doping (per Cu atom) $p = 0.14$ (underdoped $T_c = 92$ K, UD92) to $p = 0.24$ (overdoped $T_c = 47$ K, OD47), for both low temperatures (blue, T << $T_c$) and high temperatures (red, T ~ $T^*$ for pseudogap regime, T > $T_c$ for non-pseudogap regime, generally referred to as T >> $T_c$ hereafter (*24*)). The spectra in the deeply overdoped region are characterized by prominent Bogoliubov quasiparticle peaks at low temperature (T << $T_c$) and pronounced coherent peaks in the normal state (T >> $T_c$). The extracted antinodal gap size $\Delta_{AN}$ at T << $T_c$ is plotted as function of the superconducting transition temperature in Fig. 1C, revealing a remarkable 4-fold change in contrast to the only 2-fold change in $T_c$ across the doping range of interest. The gap-to-$T_c$ ratio reaches the weak coupling *d*-wave BCS value in the doping range where $T_c \lesssim 65$K (lower dashed line at $2\Delta/k_BT_c$ ~ 4.3), and the scattering dynamics can be described by Fermi-liquid theory (Fig. S2). Figure 1D shows the doping dependence of the gap-to-$T_c$ ratio $2\Delta/k_BT_c$, which rapidly increases from ~4.3 at $p$ ~ 0.22, a doping where $T_c$ ~ $T^*$ (*22, 23*), to ~10 at $p$ ~ 0.19, a doping where a pseudogap critical point has been suggested (*2, 25*). In this rapid process of the superconducting character change, $T_c$ rises from 65K to 95K despite the increasing presence of the competing pseudogap (*23*). It should be noted that throughout this doping range, the low temperature spectral gap takes the *d*-wave form, which means the antinodal gap size $\Delta_{AN}$ embodies all the *d*-component (also known as the nodal gap $\Delta_N$) in this discussion (Fig. S3) (*22, 24*). This is to be distinguished from the underdoped region, where the energy gap deviates from the *d*-wave form (*4, 22*). Indeed, $\Delta_{AN}$ and $\Delta_N$ show consistent behavior (Figs. 1, C and D).

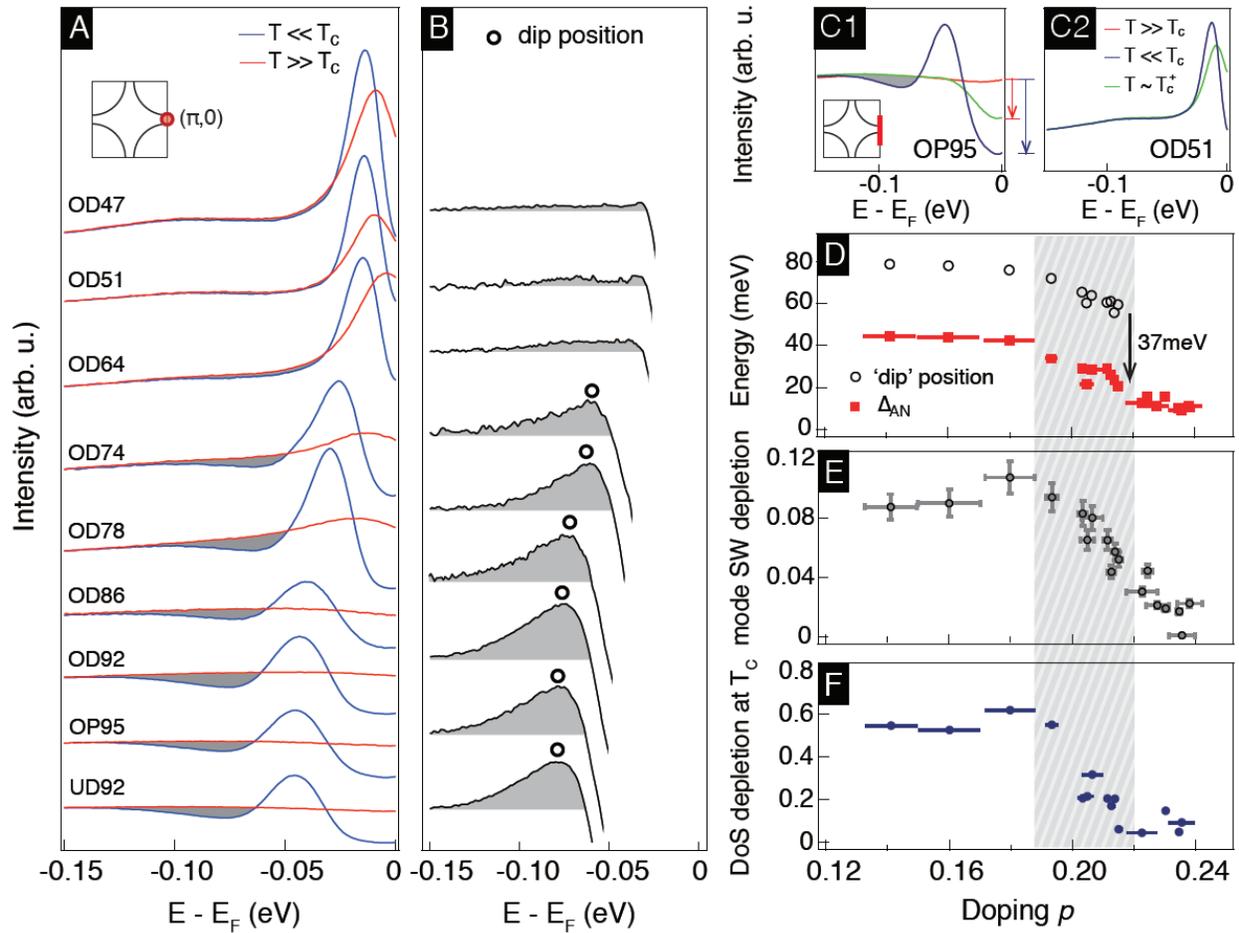

**Fig. 2. Abrupt growth of the electron-phonon coupling strength with underdoping.** (A) The $(\pi,0)$ FD-EDCs taken at low temperature (blue) and high temperature (red, $T \sim T^*$ for pseudogap regime, $T > T_c$ for sufficiently overdoped regime) from heavy overdoping (OD47) to slight

underdoping (UD92). (B) The spectral intensity difference between the two temperatures for each doping. The black circles denote the dip maxima. (C) The integrated FD-EDC over the antinodal momenta at low temperatures (T << $T_c$, blue), just above $T_c$ (T ~ $T_c^+$, green) and high temperatures (T >> $T_c$, red) for $p$ = 0.16 sample (OP95, C1) and $p$ = 0.23 sample (OD51, C2). The momentum integration range is noted by the red bar in the inset schematic BZ. The spectra are normalized to the total spectral weight between 200meV and 300meV binding energies. (D) The doping dependent dip position and the co-evolving antinodal gap $\Delta_{AN}$. (E) The doping dependent spectral weight of the dip (grey area in B) relative to the total spectral weight of the normal state. (F) The doping dependent antinodal spectral weight depletion at $T_c^+$ (red arrow in C1) relative to the full depletion at T << $T_c$ (blue arrow in C1), derived from the integrated FD-EDCs' intensity within ±2 meV of $E_F$.

With the rapid doping evolution of the superconducting character, the EPC around the Brillouin zone (BZ) boundary also shows a dramatic change. Figure 2A shows the FD-EDCs from ($\pi$,0), where the peak directly reflects the antinodal band bottom (high temperatures, red), and the spectral minima in the 'peak-dip-hump' structure are minimally disturbed by the bonding band (low temperatures, blue) (24). Such spectral dips are commonly interpreted as the fingerprint of electron-boson coupling, and become prominent with strong EPC (*16, 18, 24, 26*). This is inferred from the variation with doping of the difference between low and high temperature spectra (highlighted in grey); the spectral weight of this difference increases with growing EPC (Fig. 2B, Fig. S4). Note that the effect of the weakly temperature dependent bi-layer splitting is minimized with this analysis (*24*). As shown in Fig. 2D and Fig. S6B, the dip position tracks the rapid evolution of $\Delta_{AN}$ at T << $T_c$  by a nearly doping independent energy offset of ~37meV, which agrees well with the energy of the $B_{1g}$ oxygen bond buckling phonon mode (*18, 24*) and is qualitatively consistent with previous scanning tunneling microscope studies (*20*). Such a mode was also revealed in recent high resolution photoemission studies on similarly overdoped systems near the antinode (*27*). Surprisingly, the spectral weight associated with the dip, which reflects the EPC strength (*24*), abruptly grows between $p \sim 0.22$ and $p \sim 0.19$ (Fig. 2E). This, along with the similar behavior in doping dependent charge transport properties (*28*), cannot be explained by a simple chemical potential shift, which has yet to cross the band bottom at $(\pi,0)$ (Fig. 2A). For comparison, the pre-depletion of the density of states (DoS) from the high temperature normal state to $T_c$ (ratio between the red and blue arrows in Fig. 2C, also see Fig. S6C) is plotted in Fig. 2F. Compared to the conventionally used $T^*$ or $\Delta_{AN}$, this quantity better disentangles the pseudogap contribution from the bulk superconductivity, reflecting the direct DoS depletion effects dominated by the pseudogap. Clearly, all the three quantities – superconductivity (Fig. 2D), pseudogap (Fig. 2F) and EPC strength (Fig. 2E) – show concomitant onset and rapid change in the interval $p = 0.19 \sim 0.22$ in a highly correlated fashion.

Figure 2C highlights the dramatic contrast between the two distinct regimes – the complex strange metal regime near optimal doping (C1, strong pairing, strong EPC and pseudogap) and the behavior in the deeply overdoped region that appears to submit precisely to the classic BCS rules (C2, weaker pairing, weaker EPC and no pseudogap) – by plotting integrated spectra near the antinode. The contrast in the spectra is most striking in the normal state where no quasiparticles exist in the strange metal state, but Fermi-liquid quasiparticles are apparent in the deeply overdoped regime. The main discovery is that the strange metal phase not only involves strong correlations and incipient intertwining electronic orders but also lattice degree of freedom (in particular the $B_{1g}$ phonon for this system). The peak-dip-hump structure is the low temperature remnant of such intertwinement (after superconductivity has already set in), and is largely smeared out at high temperatures by the spectral incoherence characteristic to the strange metal state.

Combined with the doping dependent investigation (Fig. 2. D-F), we suggest that the increasing electronic correlation, marked by the entrance into the strange metal regime and the emergence of pseudogap, may have triggered this sudden EPC enhancement and the complex feedback interaction that follows. It is likely that between $p = 0.22$ and 0.19, the charge carriers slow down, the charge screening becomes inoperative, and the strange metal takes over (also see Fig. S8). With further underdoping, stronger pairing, EPC and the pseudogap become closely tied phenomena (Fig. 3 inset, top right). Although the electronically driven charge density wave does not result in significant static lattice deformation, the 'dynamical' effects of EPC are indeed abundant in the underdoped region – giant phonon anomalies (*29*), the large sensitivity of the charge order and superfluid density to isotope substitution (*19*), as well as the strengthening of the superconductivity caused by phonon pumping (*30*). Our present results indicate that the non-Fermi liquid nature of the strange metal appears to have a drastic impact on the fundamental physics of the EPC, which may well be at the origin of these highly anomalous EPC effects on the intertwined orders at low temperatures (*31*). Consequently, EPC and strange metal physics play together to enhance the system's tendency to order, inducing instability that leads to CDW or more complex electronic textures (*3*). Further underdoping leads to extreme EPC amplified by correlation effects, where the charge carriers eventually develop polaronic signatures (*13*). In particular, the $B_{1g}$ phonon is suggested to accommodate the *d*-wave form factor of both the CDW (particle-hole channel) and the superconductivity (particle-particle channel) in the cuprate systems, providing pathways for the lattice to interact with the electronic orders (*32*).

Indeed, the pseudogap temperature $T^*$ near optimal doping receives a substantial increase from single-layer to bi-layer Bi-based cuprate systems (Fig. S8C). Here only in the multi-layer case does the oxygen $B_{1g}$ mode significantly couple to electrons: as the $CuO_2^{n-}$ planes do not lie on a crystal mirror plane, a coupling can only arise from first order $c$-axis atomic displacements (Fig. 3 insets, left). Such layer dependence also holds when the tri-layer system is taken into account (*33*). Therefore with underdoping, the rapid formation of the pseudogap and enhanced EPC could be perceived as a result of a positive feedback under increasing electronic correlation, which efficiently reinforces the system's already strong tendency to order.

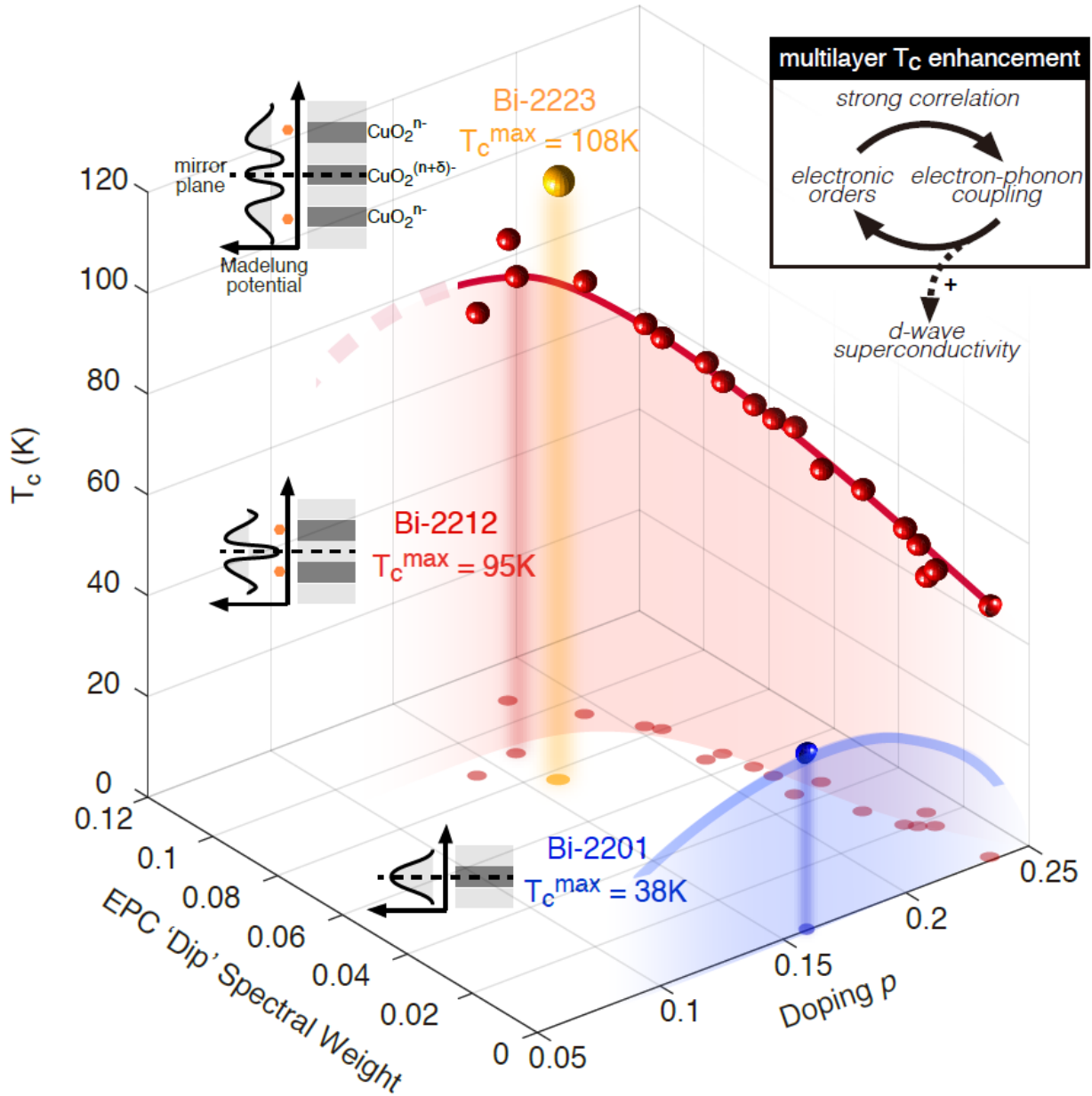

**Fig. 3. Intertwined growth of the superconductivity and the electron-phonon coupling tuned by the hole concentration.** The red line is an illustration of the $T_c$ in Bi-2212 ($T_c^{max}$ = 95 K). The blue shade and line represent the single-layer Bi-2201 system, where the coupling to the $B_{1g}$ mode is weak and $T_c^{max}$ is only 38 K. The yellow ball represents the optimally doped tri-layer Bi-2223 where $T_c^{max}$ is 108 K. The top-right inset shows the intertwined relation between the pseudogap and the EPC under strong electronic correlation. The Madelung potential and the lattice stacking

along the *c*-axis are schematically depicted for the single-layer, bi-layer and tri-layer systems. The dark grey blocks represent the $CuO_2^{n-}$ plane, and the light grey blocks represent the charge reservoir layers ($Ca^{2+}$, SrO, $BiO^+$). The orange dots mark the $CuO_2^{n-}$ planes that experience to the first order a non-zero out-of-plane electric field.

More intriguingly, superconductivity may benefit from this feedback loop. Figure 3 compiles the superconducting $T_c$ of the single-layer (Bi-2201, blue), bi-layer (Bi-2212, red) and tri-layer (Bi-2223, orange) systems as a function of doping and antinodal EPC dip strength. The maximum $T_c$ sees a significant increase from single layer to multilayer systems. Although the growing electronic correlation strength from the overdoped regime towards the optimal doping may have strengthened the pseudogap (*1 - 4*), it also joins efforts with the enhanced EPC to achieve the higher $T_c$ in multi-layer cuprate superconductors than in their single-layer counterparts (*33*). As a corroborative example, a recent photoemission study on monolayer FeSe film system demonstrated the experimental viability to have a secondary phononic channel as the superconductivity 'enhancer' (*16*), boosting $T_c$ by as much as 50% in comparison to the uncoupled case. The oxygen bond-buckling $B_{1g}$ phonon couples strongly in the multilayer cuprates, thanks to the steep Madelung potential going through the $CuO_2^{n-}$ planes that do not constitute the crystal mirror plane (orange markers, Fig. 3 inset). This phonon mode is also known to favor small-*q* scattering thanks to its preferential coupling to electrons in the antinodal region and is inherently poised for *d*-wave superconductivity enhancement (*15*). This complex involvement of EPC in the pairing process may also be the underlying reason for the dynamical and diverse manifestation of isotope effects (*19 - 20*). Therefore, it may be insufficient to regard Cooper pairing as driven by pure electron-electron interactions in the optimal and underdoped regions.

On the other hand, the deeply overdoped regime emerges as a weakly coupled 'simpler' superconductor with a respectably high $T_c$ in excess of 50K, which offers a cleaner platform to investigate the pairing mechanism. Meanwhile with enhanced correlation in the strange metal regime, EPC and the intertwined electronic orders form a positive feedback loop (Fig. 3 inset, solid arrows) that ultimately provides an additional pathway for $T_c$ enhancement by EPC in the multilayer systems (dashed arrow). From a broader perspective, cuprate superconductors, given this interpretation of their rich phase diagram, may be generalized as a new model system for testing EPC enhanced superconductivity through multiple underlying channels. Systematic investigation of the $B_{1g}$ phonon self energy around the critical doping via various scattering techniques may provide valuable insights into such scenarios.

**Acknowledgments:**

The authors wish to thank R. Hackl, S. Kivelson, S.-L. Yang, M. Yi, Y. Wang, and E. W. Huang for discussions. ARPES experiments were performed at Stanford Synchrotron Radiation Lightsource, SLAC National Accelerator Laboratory, and at Stanford Geballe Laboratory for Advanced Materials at Stanford University. This study is supported by the U.S. Department of Energy, Office of Science, Office of Basic Energy Sciences under Contract No. DE-AC02-76SF00515. D.-H.L. was supported by the U.S. Department of Energy, Office of Science, Basic Energy Sciences, Materials Sciences and Engineering Division, grant DE-AC02-05CH11231.

# Supplementary Materials

**Materials and Methods**

<u>Sample characterization</u>

  To quantify the surface oxygen variation via typical *in-situ* cleaving process involved in ARPES measurement, the surface $T_c$ is determined based on the in-gap DoS depletion (Fig. S1). A comparison of bulk and surface $T_c$'s for near optimally doped samples, whose oxygen levels are close to the as-grown condition and are more stable upon *in-situ* cleaving in vacuum, are also listed in Table. S1.

  Figure S1A plots the temperature dependent in-gap spectral weight integrated within 4 meV of $E_F$ for all the doping at the antinode. $T_c$ is extracted from the steepest drop on the curve. Fig. S1B and S1C show the process for OD51 and OP95 samples. The $T_c$ uncertainty is reflected by the temperature step size at the steepest drop on the curve. For samples near the optimal doping where oxygen loss is less severe than the deeply overdoped region, the bulk $T_c$ shows good agreement with the *in-situ* surface $T_c$ derived with the aforementioned method. For extremely overdoped samples (bulk $T_c$ at 55 K and 50 K), various degree of oxygen loss takes place during the sample heating and the cleaving process, which mostly yield higher *in-situ* $T_c$ than the bulk $T_c$ before the treatment. Systematic doping dependent photoemission studies by Kondo et al (*34*) also consistently show agreement between the sample $T_c$ and the steepest DoS drop temperature. It should be noted that such deeply overdoped regime has rarely been accessed for Bi-2212 system in the past, and is crucial to understand the weak coupling properties of its superconductivity.

  The effective hole doping *p* is determined following the empirical parabolic relation (*35*):

$$T_c/T_c^{max} = 1 - 82.6*(0.16 - p)^2$$

Where $T_c^{max}$ is 95 K. The doping uncertainty is propagated from the $T_c$ uncertainty according to this parabolic relation. The measurement temperatures – low temperature ($T \ll T_c$, base temperature), intermediate temperature ($T_c^+$, just above $T_c$), and high temperature ($T \sim T^*$ for the pseudogap regime, $T > T_c$ for sufficiently overdoped regime) are also listed in Table. S1 according to the definition in the main context.

<u>ARPES measurements</u>

The $Bi_2Sr_2CaCu_2O_{8+\delta}$ single crystals were grown with optical floating zone method. They were cleaved *in-situ* under ultrahigh vacuum. ARPES measurements were performed with pressure lower than $5x10^{-11}$ torr at BL5-4 of Stanford Synchrotron Radiation Laboratory with 9.0, 18.4 and 22.7eV photon energies. The bulk $T_c$'s and effective doping level as measured by the magnetic susceptibility are listed in Table. S1. The Pb doping is introduced to suppress the Bi-O layer superstructure along $(\pi,\pi)$ direction.

**Supplementary Text**

<u>Fermi liquid – non-Fermi liquid self-energy</u>

It is observed that when more holes are introduced into the Bi-2212 system, the normal state gradually crosses over to a Fermi liquid in terms of its electronic self-energy *(36, 37)*. Given the overlapping superstructures and shallow band dispersion near the antinode region, we turn to the nodal spectrum for the self-energy extraction. The low temperature (~10 K) imaginary part of the self-energy at nodal $k_F$ is plotted in Fig. S2A for $p = 0.06$ ($T_c = 22$ K), Fig.S2B for $p = 0.11$ ($T_c = 75$ K) and Fig.S2C for $p = 0.23$ ($T_c = 56$ K). In a Fermi liquid system, the electron-electron interaction induced scattering rate should take a quadratic form in terms of the binding energy $\omega$

= |E - E_F| when ω << E_bandwidth. Instead, it was observed to follow a linear-ω dependence around optimally doped and underdoped region in high $T_c$ cuprate compounds (*38*). Early work in extremely overdoped Bi-2212 indicated $\omega^2$-dependence at low temperature and the crossover to linear regime at high temperature (*36, 37*).

With improved statistics and resolution, we decompose $\Sigma''$ into the impurity scattering channel (constant $\Sigma''_{imp}$), the main electron-phonon coupling channel ($\Sigma''_{ep} \sim erf(\omega\text{-}\omega_{ph})$) and the electron-electron interaction channel ($\Sigma''_{ee} \sim \omega^n$). The fitted index $n$ is plotted in Fig. S2D as a function of temperature. In OD56 samples, nearly quadratic energy dependence is observed for $\Sigma''_{ee}$. This serves as a spectroscopic indication of Fermi liquid behavior for the extremely overdoped region when the superconductivity is absent, utterly deviating from the nearly linear behavior in deeply underdoped and near optimally doped samples. Therefore at least near the nodal region, the system becomes Fermi-liquid-like at sufficient overdoping.

Determining *d*-wave superconducting gap

It should be noted that in very underdoped Bi-2212 ($p < 0.12$) that the momentum dependence of the low temperature energy gap deviates from the simple *d*-wave form near the antinode (*22, 39*) where the pseudogap is strong. In such regime, the *d*-wave component has been extracted from the near nodal region based on a linear extrapolation under |cos($k_x$) - cos($k_y$)|/2 scaling (*22*). However, the momentum dependence of the low temperature energy gap recovers the *d*-wave form when $p > 0.13$ in Bi-2212. Therefore in the doping range investigated by the current work ($0.14 < p < 0.24$), the *d*-wave component of the gap, commonly known as the 'nodal gap' $\Delta_N$ or 'gap velocity' $v_\Delta$, is almost identical in value to the *d*-wave energy gap size at the antinode $\Delta_{AN}$. The visual appearance of an 'arc-like' feature in the deeply overdoped regime is largely due to the finite resolution effect combined with the small energy gap close to the node (*22, 34*). Figure S3 explicitly confirms such observation within experimental resolution, for both the extremely overdoped case (OD56, $p = 0.23$) and the near optimally doped case (OD86 $p = 0.19$).

Given the $d$-wave gap, the $d$-wave BCS gap-to-$T_c$ ratio, and the parabolic nodal imaginary electronic self-energy, the deeply overdoped Bi-2212 system presents itself as a weakly-coupled $d$-wave high $T_c$ BCS superconductor, where electronic interactions plays an important role. While the underlying pairing mechanism remains an unsolved mystery, this system presents a clean testbed. The lack of a spectral 'dip' does not rule out spin fluctuation mediated pairing, as in the weak coupling regime the pairing boson need not be visible in the tunneling or ARPES spectra. However, approaching the van Hove singularity with overdoing, most sign-flipping $(\pi,\pi)$-scattering based bosonic pairing naively should see a dramatic enhancement in the $d$-wave channel, which has not been clearly observed. Taken as a whole, this suggests the need to consider other pairing mechanisms on the full energy spectrum. The strong material dependence in $T_c$, even at this high doping, implies a potentially important role played by the lattice.

Temperature dependent electron-phonon coupling simulation

The bi-layer splitting in Bi-2212 historically added complexity to the discussion of the peak-dip-hump structure (*40, 41*). This inter-layer interaction induced bonding-antibonding band splitting may also contribute to this dip feature at $(\pi,0)$, which should be only weakly temperature dependent and reduces with underdoping. In this study, the main focus is on the lower energy antibonding band, which contributes significantly more density of states near the Fermi level. To minimize the contribution from the bonding band and superstructure structure effects, the spectral weight difference between the high temperature and low temperature FD-EDCs is employed to extract the highly temperature dependent mode coupling strength. Fig. S4 shows the simulated antinodal spectra with both the 35 meV $B_{1g}$ mode ($g^2 = 0.078$ eV$^2$) and the 75 meV half-breathing mode ($g^2 = 0.032$ eV$^2$). The band structure is determined by the following tight-binding fit to capture the Fermi surface shape

$$\epsilon_k = -2.52 \left( cos k_x a + cos k_y a \right) + 2.12 \, cos(k_x a) \, cos(k_y a) \qquad (S1)$$

and a momentum dependent band renormalization to capture the highly anisotropic band shape and velocity

$$\lambda_{ee} = 1 + 14 \, sin^6 \left( \sqrt{\frac{k_x^2 + k_y^2}{8}} a \right) \qquad (S2)$$

Imaginary part of the EPC self-energy is calculated from single iteration Eliashberg framework (*15*), and the real part is obtained by performing KK transform on the imaginary part. The momentum dependence of the two modes are described by the following functional forms (*15*)

$$g^{hbr}(k, q) = g_0^{hbr} \sqrt{sin^2 \frac{q_x a}{2} + sin^2 \frac{q_y a}{2}} \qquad (S3)$$

$$g^{B1g}(k, q) = g_0^{B1g} \left( sin \frac{k_x a}{2} sin \frac{k_x' a}{2} cos \frac{q_y a}{2} - sin \frac{k_y a}{2} sin \frac{k_y' a}{2} cos \frac{q_x a}{2} \right) \qquad (S4)$$

where $k$ is electron initial state momentum, $k'$ is the electron final state momentum, $q$ is the phonon momentum, and $a$ is the in-plane lattice constant.

The $q$-summed EPC constant $\lambda_{k=(\pi,0)}$ for the half-breathing mode and the $B_{1g}$ mode are 0.09 and 0.36 respectively at 170 K, which evolve to 0.06 and 0.41 at 30 K. The maximum total real part of the self-energy rises from ~20 meV at 170 K to ~40 meV at 30 K. It should be noted that the superconductivity's enhancement effect on the $B_{1g}$ mode is reflected in the simulation via both the increased EPC constant as well as the gap-shift behavior at low temperatures.

Figure S4(C) plots the $(\pi,0)$ EDCs at 170 K (red) and 30 K (blue), and their difference (black). Thermal broadening effect is taken into account by incorporating the Fermi liquid self-energy. It's clear that both the $B_{1g}$ mode and the half-breathing mode are identified in the difference curve, where the black circle marks the superconducting gap-shifted mode position. Similar simulation is also performed with the EPC strength reduced to 10% of the original value, and are listed in Fig. S4D-F. It shows that with much reduced EPC strength, the dip spectral weight also dramatically shrinks while its extrema position remains unchanged.

Figure S5 employs the same framework to address the robustness of this method against realistic experimental imperfections and interferences on the spectroscopic signature of $B_{1g}$ phonon. It can be seen that inclusion of significant energy broadening (Fig.S5B), momentum broadening and sample mosaicity (Fig.S5C), thermal broadening (Fig.S5D) and the inclusion of an adjacent bonding band (Fig.S5E) does not qualitatively change the result (Fig.S5F). It should be noted that the 'dip' spectral weight evaluated in the presence of a bonding band serves as an underestimate of the actual EPC strength. With the bi-layer splitting energy rather doping independent, such effect will not change the doping dependence reported in the current work.

Doping and temperature dependent antinodal spectra

Figure S6 lists all the doping dependent FD-EDCs taken at high and low temperatures for various momenta. $(\pi,0)$ is chosen for the spectral dip analysis, because the band bottom is in utmost proximity to the mode energy (to maximize the observable effect), directly comparable with experimental and theoretical literature results (*33, 42 - 44*), as well as experiences the least spectral contamination from the bonding band. Antinodal $k_F$ is chosen for the superconducting gap analysis because the superconducting gap is defined at $k_F$. The integrated spectra has significantly higher signal-to-noise ratio and integrates out the superstructure effect near the antinode, and is used as supporting evidence for the spectral dip doping evolution. It should be pointed out that the doping evolution of the spectral dip is consistent across all three panels.

<u>Antinodal spectral lineshape simulation with competing orders</u>

Numerical simulations based on given EPC coupling kernel $g$ (valued at 0.05, 0.15, 0.25, 0.35 in horizontal arrangement), *a priori* order parameters of the *d*-wave superconductivity ($\Delta_{dSC} = 36$ meV) and the pseudogap (represented by a short range charge density wave, whose order parameter is valued at 0, 6, 26, 46, 66, 86, 106 meV in vertical arrangement) are shown in Fig. S7(A-D). The detailed description of the model used here can be found in reference (*23*). It is shown that simply tuning either EPC or pseudogap strength is not sufficient to capture the antinodal EDC lineshape evolution from the optimal doping (large high energy spectral hump and reduced coherent peak) to the deeply overdoped regime (much weakened spectral hump and strong coherent peak). And the doping dependence is characterized by the simultaneous reduction of the quasiparticle coherence near the Fermi energy (mainly due to the pseudogap) and the enhancement of the spectral hump and dip (due to the electron phonon coupling and the pseudogap) when the system approaches the half filling. Pronounced 'dip' feature is also universally observed in multilayer cuprate superconducting systems (*33, 43, 44*). It should be noted that in the simulation, only antibonding band is considered, and the bonding band induced spectral hump is of a different nature from the correlation effect induced spectral hump discussed here.

<u>Estimation of screening properties and doping dependent T$^*$</u>

To approximate the effective screening length as function of doping, the antinodal density of states (per unit cell) from the antibonding band prior to the superconducting transition is calculated based on the product of the normal state band dispersion (at ~ T$^*$) and the relative spectral weight depletion $d_{PG}$ from T$^*$ to T$_c^+$ (Fig. 2F), as indicated by Eqn. S5.

$$DoS_{AN} = 2\,a^2\,\frac{|k_F|^2}{v_F \cdot k_F}\,(1 - d_{PG}) \qquad\qquad (S5)$$

where $a$ = 3.82 Å is the tetragonal in-plane lattice constant, $k_F$ is the antinodal Fermi momentum, $v_F$ is the Fermi velocity at the antinode, and $d_{PG}$ is the relative spectral weight depletion from $T^*$ to $T_c^+$ as plotted in Fig. 2F. The result is the effective antinodal density of states shown in Fig. S8A. The corresponding Thomas-Fermi screening length is evaluated according to Eqn. S6, and is plotted in Fig. S8B. Between $p$ = 0.19 - 0.22, $r_{TF}$ crosses half Cu-O bond length.

$$r_{TF} = \sqrt{\frac{\varepsilon_0}{e^2\,DoS_{AN}}} \qquad\qquad (S6)$$

Including the combined effect from band renormalization and the pseudogap gives rise to a rapidly changing Thomas-Fermi screening length across the doping range of interest. However, such a treatment of screening remains overly simplistic. First, Thomas-Fermi screening is based on a static charge response, which does not capture the finite energy spectral weight transfer in the strange metal state. Second, the screening of any specific phonon should depend on the detailed form of its coupling (*15*). Additionally, the assumption that the antinodal low energy electrons, which are most affected by the pseudogap, dominate screening need not remain true for all phonons.

These considerations suggest that the sudden enhancement of the EPC may go well beyond simple screening. For example, the $B_{1g}$ phonon, with a $d$-form factor to its out-of-plane oxygen displacements, naturally couples to $d$-form factor CDW fluctuations below the critical doping observed in STM (*45*). Thus there is a good reason to expect that below the critical doping the CDW becomes intertwined with the phonon degree of freedom, strengthening the EPC. In general the weakening of the EPC in the overdoped regime is due to the weakening and loss of the "intertwined orders" and the disappearance of the strange metal state.

Figure S8C plots the pseudogap closing temperature $T^*$ as function of hole doping for single-layer and bi-layer Bi-based cuprates, determined from ARPES (*22, 23, 34, 46, 47*) measurements conducted with similar conditions. It's demonstrated that multi-layer systems consistently exhibit higher $T^*$ than their single-layer counterparts near the optimal doping, where in the latter electrons hardly couple to the $B_{1g}$ Cu-O bond buckling mode due to CuO$_2$ plane being the crystallographic mirror plane. It should also be pointed out that, in the sufficiently overdoped region ($p > 0.19$), $T^*$ starts to merge between Bi-2212 and Bi-2201, consistent with our observation of the unique presence of strong EPC enhancement in near-optimally doped bi-layer systems. This is also supported by a recent systematic Nernst coefficient study in YBCO and LSCO (*48*).

<u>Direct comparison of the layer dependent FD-EDCs at the zone boundary</u>

Figure S9 lists the high and low temperature ($\pi$,0) FD-EDCs for Bi-2223 ($T_c$ = 108 K), optimally doped Bi-2212 ($T_c$ = 98 K and overdoped $T_c$ = 51 K), and Bi-2201 ($T_c$ = 38 K). It is clearly seen that the spectral dip is unique in the multi-layer system, and their strength are comparable. The reason for the tri-layer system not exhibiting a stronger 'dip' than the bi-layer system is possibly that the middle CuO$_2$ layer sits on the crystalline mirror plane, therefore will not effectively couple to the $B_{1g}$ mode. The optimally doped Bi-2201 system still sees strong coexistence of the pseudogap and superconductivity, but differs from the Bi-2212 system also by the absence of superconducting quasiparticle peaks. Regardless, the mode coupling signature is noticeably missing in the single-layer system.

<u>Comparison with Raman, inelastic neutron and X-ray scattering spectroscopy</u>

Given the sudden change of EPC strength revealed by the electronic spectral function, comparison with phonon self energies in this doping range shall provide additional insights.

Raman scattering in the $B_{1g}$ channel does show qualitative spectral change between underdoped and deeply overdoped Bi-2212 (*49*). However, the phonon lineshape in Raman response is controlled not only by the electron-phonon coupling, but also the charge carrier properties (*49*). Extraction of the electron-phonon coupling component is thus complicated by the highly doping dependent electronic structure, including the influence from strong correlation effects (*50*). More importantly, typical optics and Raman measurements probe only the phonon self energy close to $q = 0$, while the electron self energy extracted via ARPES reflects the contribution from all phonon momenta. In fact, neutron scattering finds that the $B_{1g}$ phonon has a stronger coupling effect at a small but finite $q$ than from $q = 0$ (*51*). With the improving energy resolution of resonant inelastic x-ray scattering technique, direct determination of the doping-and-momentum dependent electron-phonon coupling vertex may finally become possible.

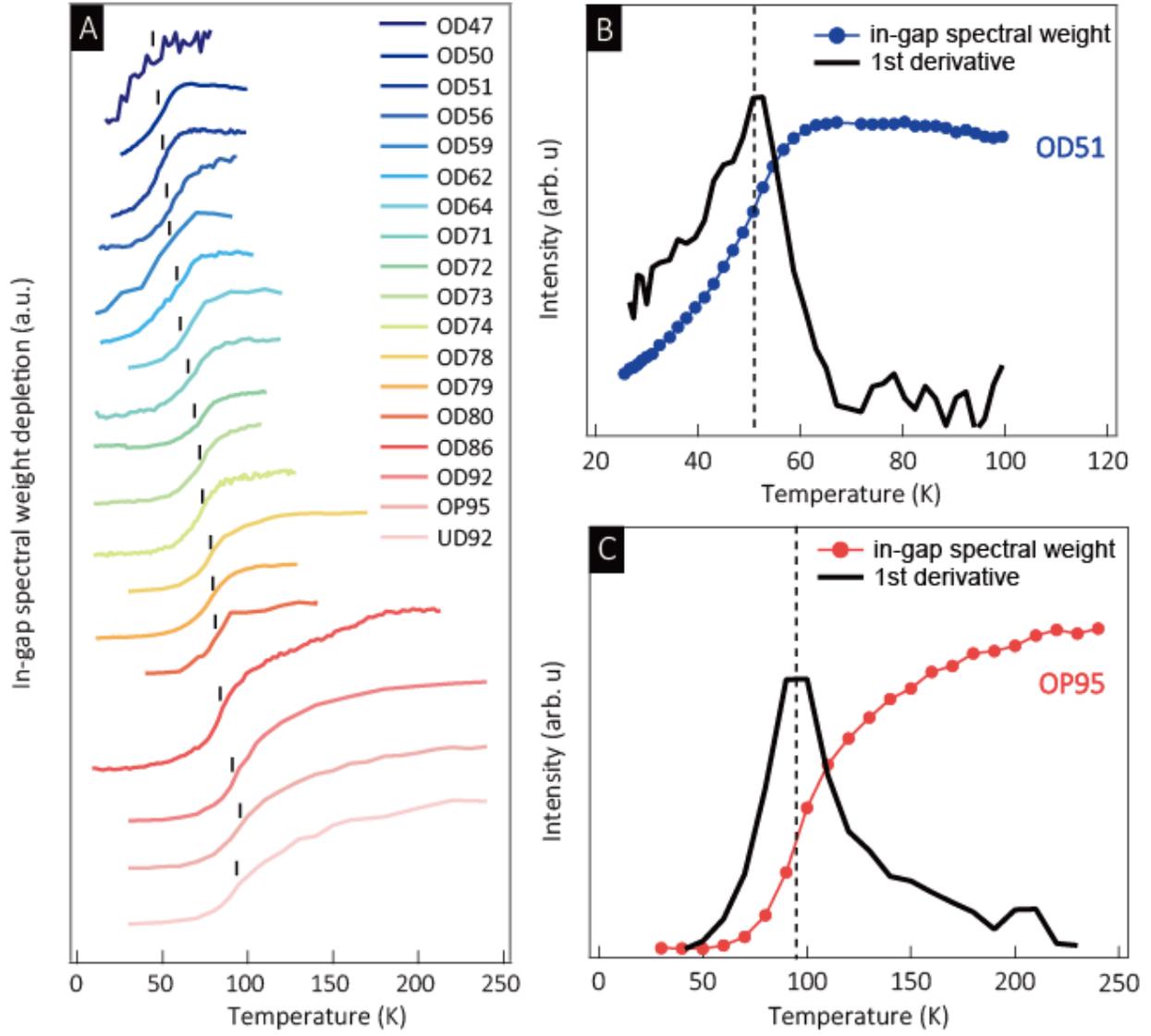

**Fig. S1**.

Determination of $T_c$ (*in-situ*). (A) List of temperature dependent in-gap spectral weight for all the dopings, integrated within 4 meV energy window of $E_F$ at the antinode. (B) The temperature dependent in-gap spectral weight (blue circle) and its temperature derivative (black line) to determine the *in-situ* $T_c$ of 51 K. (C) Similar process on OP95 sample.

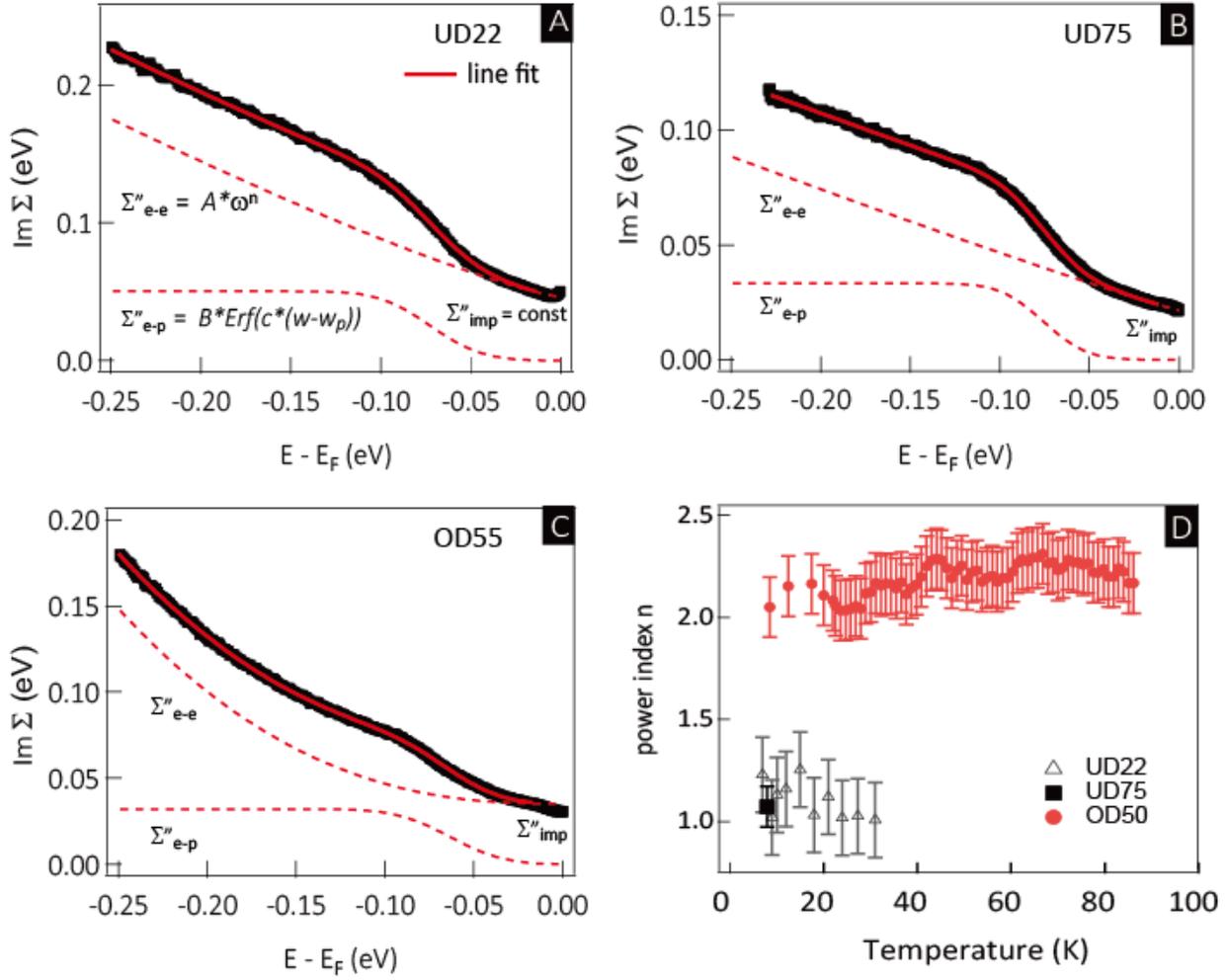

**Fig. S2.**

Crossover from non-Fermi liquid to Fermi liquid near the nodal region with increasing hole doping. (A-C) Imaginary part of the electron self-energy $\Sigma''$ for UD22, UD75 and OD56 samples. The solid red lines are fits to the data (black circle). The red dashed lines are decomposed fitting consisting of an impurity scattering term (constant $\Sigma''_{imp}$), an electron-phonon coupling term ($\Sigma''_{ep} \sim erf(\omega - \omega_{ph})$) and the electron-electron interaction term ($\Sigma''_{ee} \sim \omega^n$). (D) The temperature dependence of the power index $n$ in the electron-electron interaction term for the three dopings.

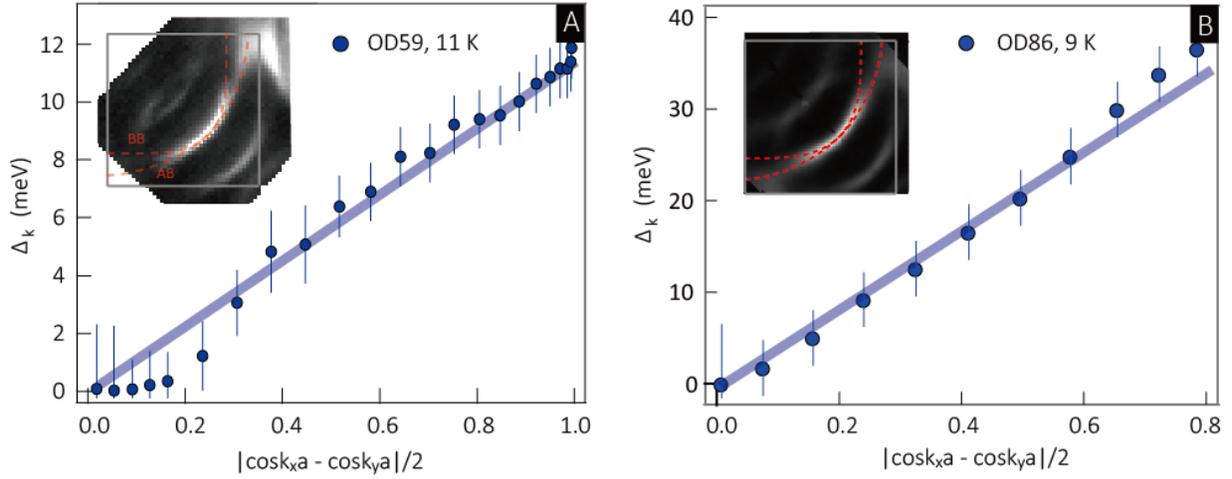

**Fig. S3.**

Linear *d*-wave gap form at low temperature under the |cos($k_x$a) - cos($k_y$a)|/2 scaling. The energy gap momentum dependence in OD59 (*p* = 0.23) sample at 11 K (A) and OD86 (*p* = 0.19) sample at 9 K (B). The insets show the Fermi surface maps (integrated within $E_F$ ± 10 meV) of the two samples respectively. The red dashed lines indicate the loci of the bonding band (BB) and the antibonding band (AB).

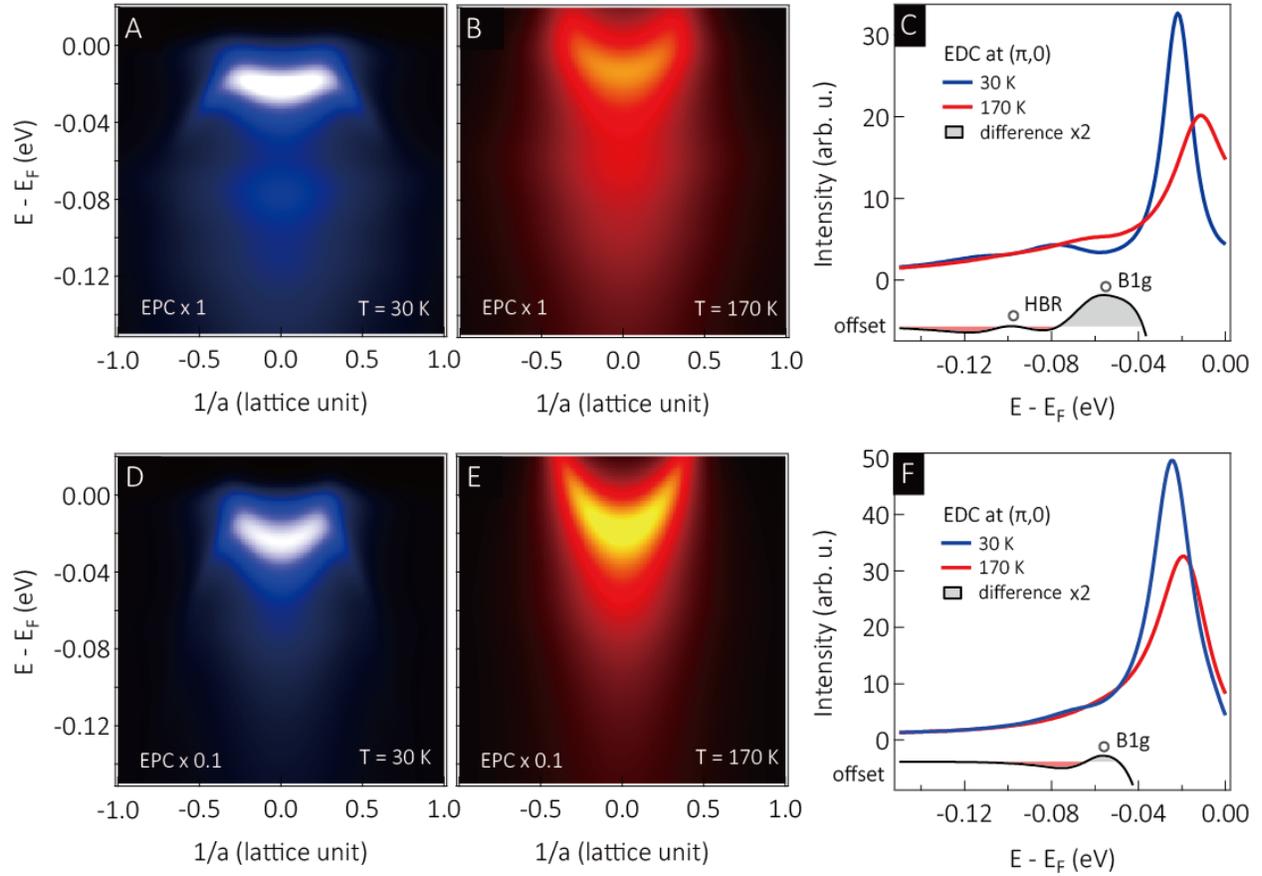

**Fig. S4.**

Simulated antinodal spectra with moderate EPC to both the half-breathing mode and the $B_{1g}$ buckling mode. (A) Superconducting state at 30 K. (B) Normal state at 170 K. (C) High temperature and low temperature EDCs and their intensity difference (black) at ($\pi$,0). (D-F) Similar calculation to (A-C) but with the phonon self-energies reduced to 10% of before.

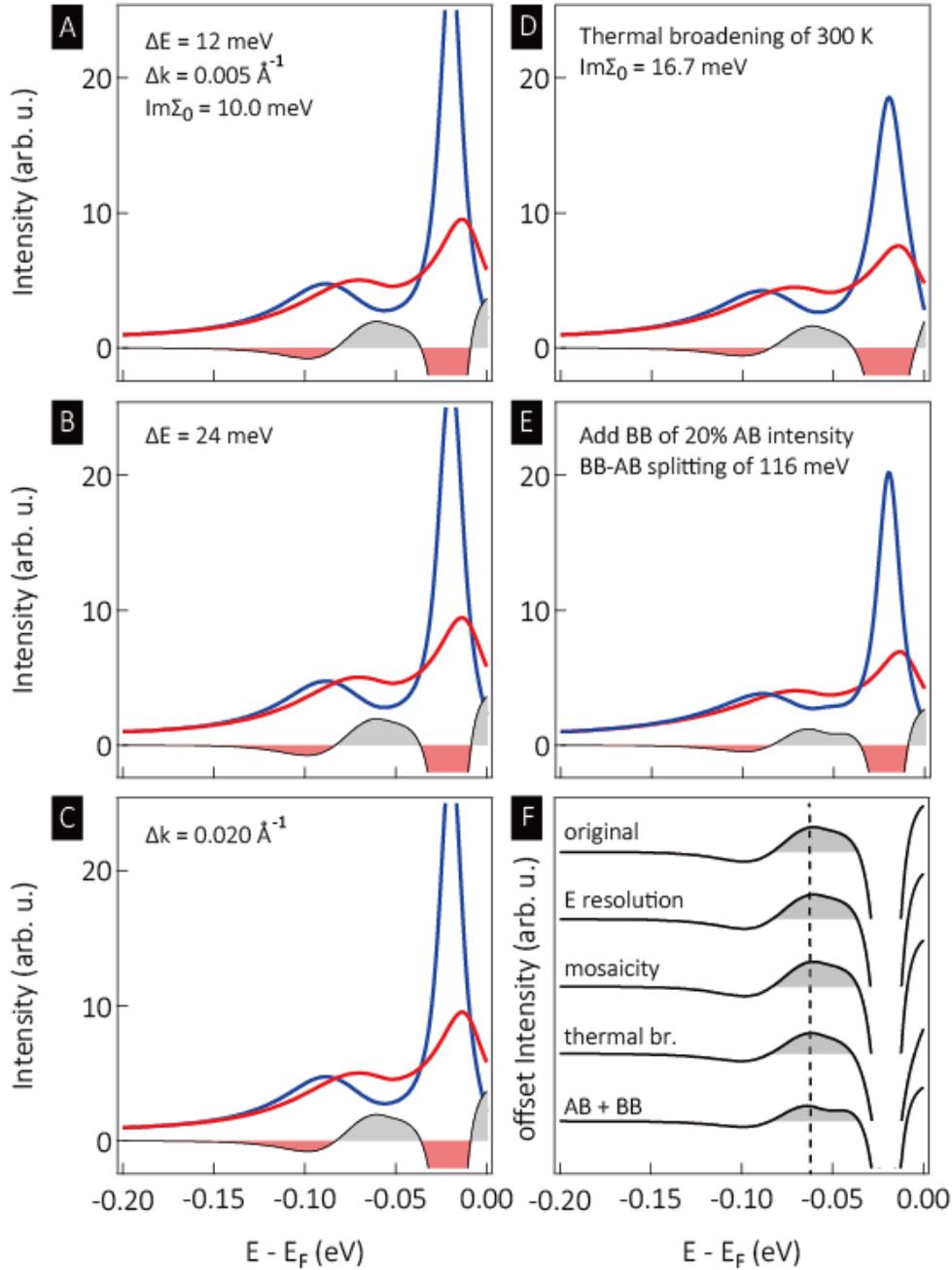

**Fig. S5.**

Simulated (π,0) spectra with moderate EPC to the $B_{1g}$ buckling mode. Shaded region – difference between red and blue curves. (A) Control EDC with energy resolution of 12 meV, momentum resolution of 0.005 Å⁻¹, constant broadening of 10 meV with a pedagogical EPC $\lambda_{B1g}$ of 0.98 at 30K (blue) and 170K (red), where the intensity is normalized to 1 at 200 meV binding energy. (B) Energy resolution effect - worsens to 24 meV. (C) Momentum mosaicity effect – momentum resolution worsens to 0.020 Å$^{-1}$. (D) Thermal broadening effect – additional constant broadening of 6.7 meV with Fermi liquid form of $(\pi k_B * 300 \text{ K})^2$. (E) Bilayer splitting effect – add in a bonding band (BB) that is split 116 meV below the antibonding band (AB) at $(\pi,0)$, where BB takes 20% intensity of AB. (F) Comparison of all high temperature – low temperature difference curves.

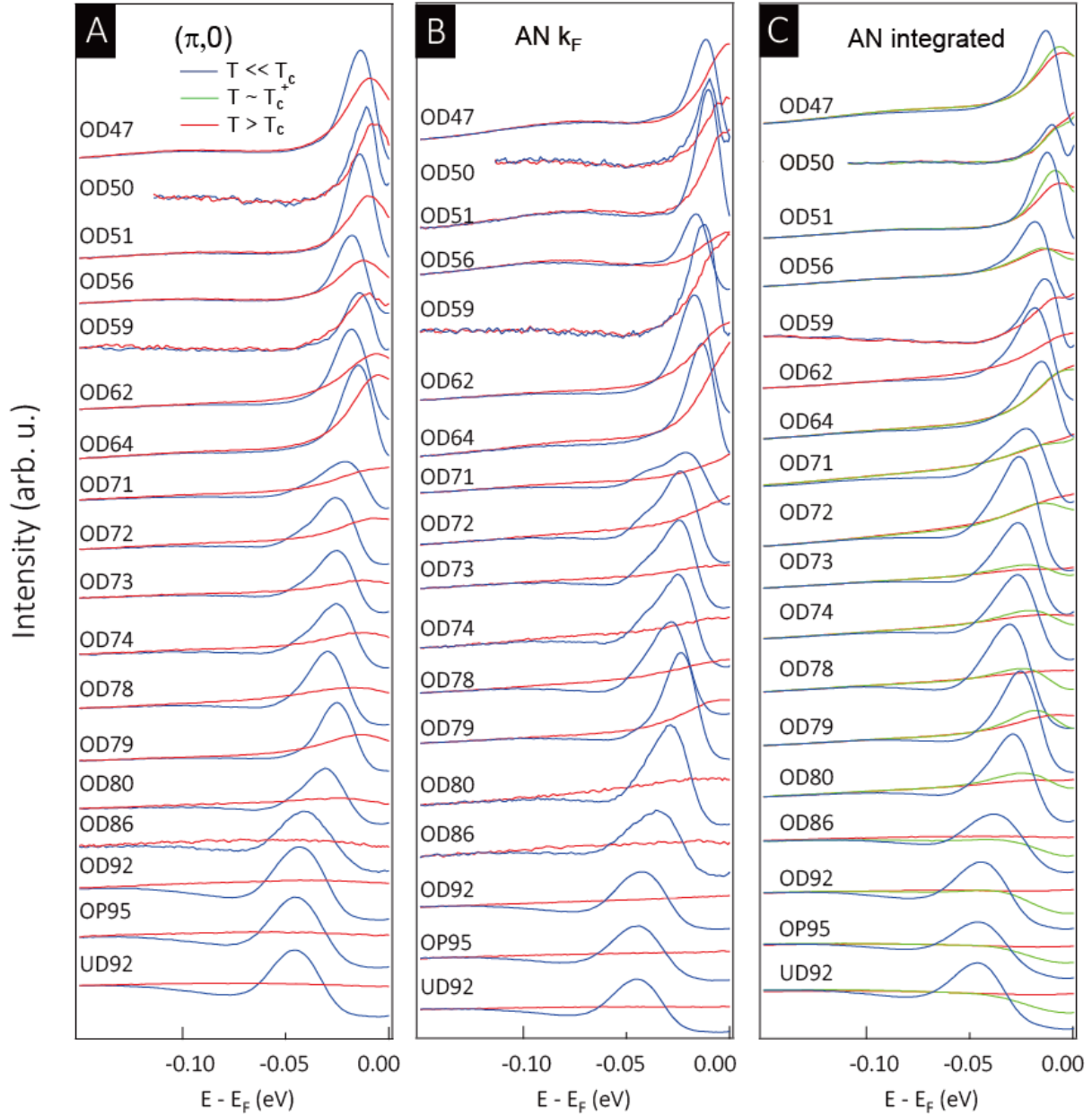

**Fig. S6.**

FD- EDCs for all dopings at (A) $(\pi,0)$, (B) antinodal $k_F$, (C) integrated along the zone boundary. Blue – low temperature in the superconducting state ($T \ll T_c$). Green – intermediate temperature right above $T_c$ ($T \sim T_c^+$). Red – high temperature in the normal state ($T \sim T^*$ for pseudogap regime, $T > T_c$ for sufficiently overdoped regime).

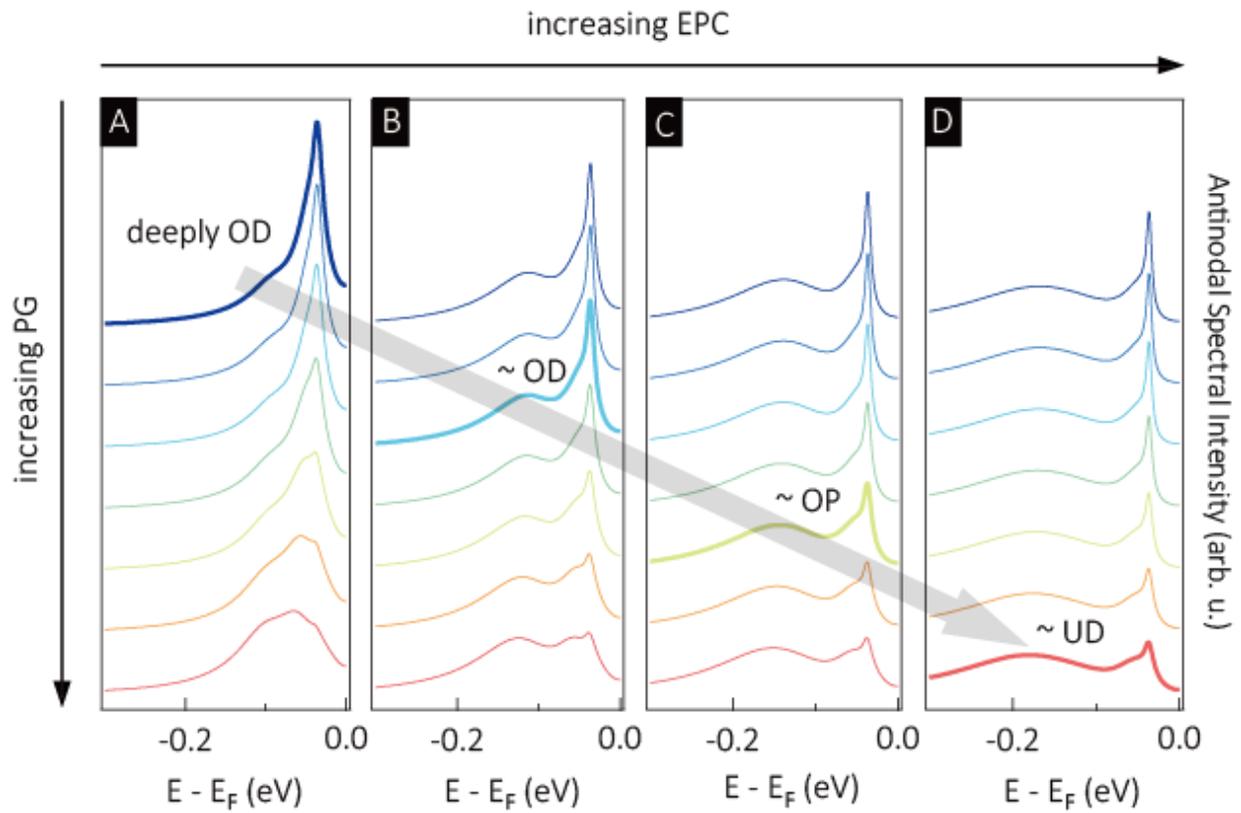

**Fig. S7.**

Numerical simulation of the combined effect of EPC and the pseudogap on the superconductivity. (A-D) Simulation of the antinodal EDC with varying mode-coupling and pseudogap strength. The thick lines are situations that resemble the experimental observations from deeply overdoped (OD) region, to the underdoped (UD) region.

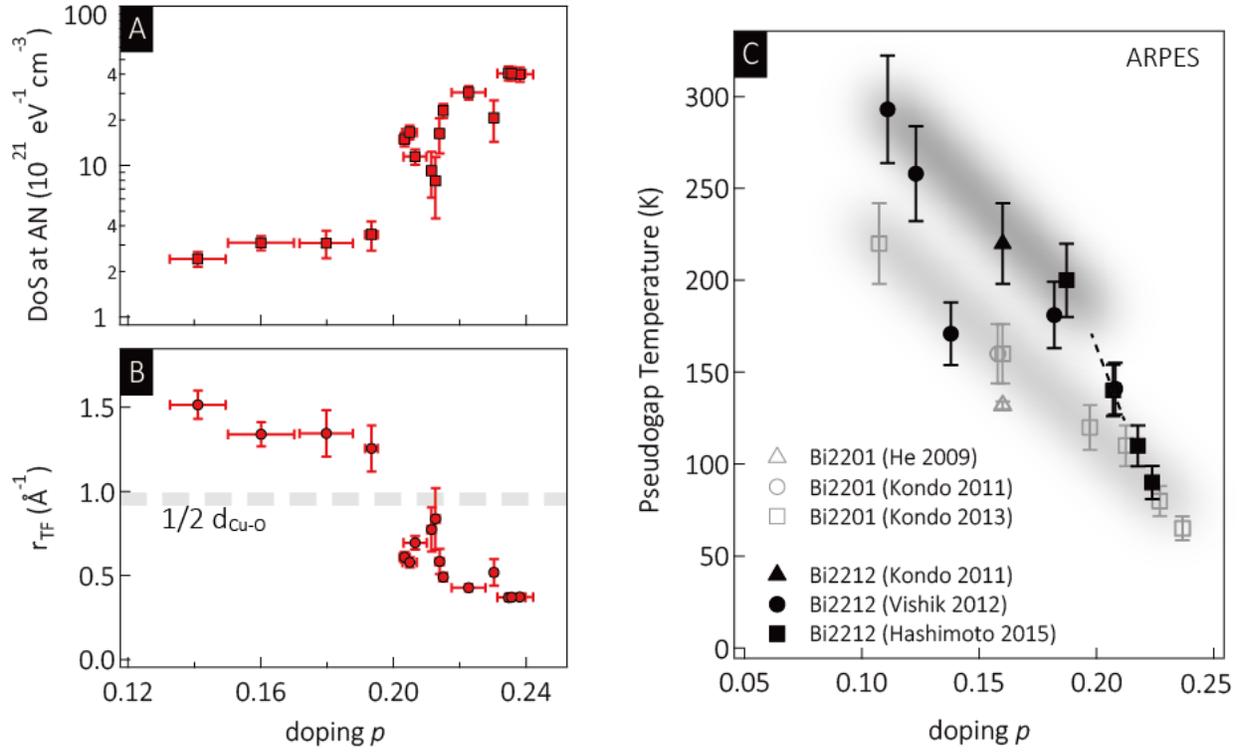

**Fig. S8.**

Experimental evidence for the interplay between EPC and the pseudogap. (A) The doping dependent electron density of states near the antinodal region prior to superconducting transition at $T_c^+$. (B) The corresponding Thomas-Fermi screening length as a function of doping. (C) Pseudogap temperature $T^*$ measured by high resolution ARPES. Grey shades are guides to the eye, where lighter color refers to single layer compounds.

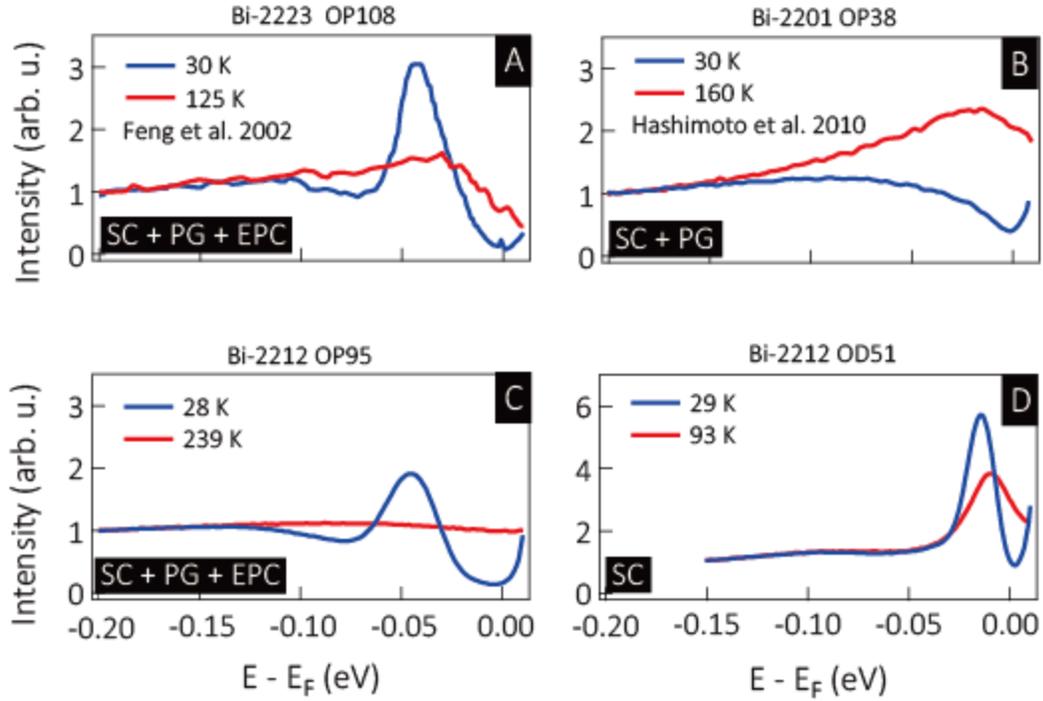

**Fig. S9.**

FD- EDCs at $(\pi,0)$ for the layer dependence. (A) Optimally to slightly overdoped Bi-2223 with $T_c$ = 108 K. (B) Optimally doped Bi-2201 with $T_c$ = 38 K. (C) Optimally doped Bi-2212 with $T_c$ = 95 K. (D) Deeply overdoped Bi-2212 with $T_c$ = 51 K. SC – superconductivity. PG – pseudogap. EPC – electron-phonon coupling.

| Sample | $T_c$ (*in-situ*) | $T_c$ (*bulk*) | hole doping | Pb-doping | $T << T_c$ | $T_c^+$ | $T \sim T^*$ |
|--------|-------------------|----------------|-------------|-----------|------------|---------|--------------|
| UD92 | $92 \pm 2.5$ | 92 | 0.141 | Y | 29 | 102 | 239 |
| OP95 | $95 \pm 5$ | 98 | 0.160 | Y | 28 | 105 | 239 |
| OD92 | $92 \pm 2.5$ | 92 | 0.180 | Y | 28 | 98 | 238 |
| OD86 | $86 \pm 1$ | 86 | 0.193 | N | 18 | 93 | 204 |
| OD80 | $80 \pm 1$ | 81 | 0.203 | Y | 15 | 88 | 124 |
| OD79 | $79 \pm 1.5$ | 80 | 0.205 | Y | 33 | 85 | 133 |
| OD78 | $78 \pm 2.5$ | 71 | 0.207 | Y | 30 | 87 | 170 |
| OD74 | $74 \pm 1$ | 55 | 0.211 | N | 15 | 85 | 145 |
| OD73 | $73 \pm 1$ | 55 | 0.213 | N | 15 | 85 | 124 |
| OD72 | $72 \pm 1$ | 50 | 0.214 | N | 15 | 85 | 103 |
| OD71 | $71 \pm 1$ | 50 | 0.215 | N | 20 | 90 | 110 |
| OD64 | $64 \pm 1$ | 50 | 0.223 | N | 22 | 76 | 94 |
| OD62 | $62 \pm 1.5$ | 65 | 0.224 | Y | 28 | 67 | 119 |
| OD59 | $59 \pm 4$ | 50 | 0.228 | N | 18 | 64 | 80 |
| OD56 | $56 \pm 1$ | 50 | 0.230 | N | 21 | 72 | 86 |
| OD51 | $51 \pm 1$ | 50 | 0.235 | N | 29 | 58 | 93 |
| OD50 | $50 \pm 5$ | 50 | 0.236 | N | 23 | 54 | 71 |
| OD47 | $47 \pm 5$ | 50 | 0.238 | N | 26 | 64 | 95 |

**Table S1.**

Sample bulk $T_c$ determined by magnetic susceptibility measurement and the *in-situ* $T_c$ determined by the in-gap spectral weight. The effective hole doping *p*, Pb presence and measurement temperatures are also listed. The experiment temperatures (low, intermediate and high) are listed in the last three columns.